# Computational reconstruction of mitochondria-encoded mammal ancestral proteins

Bohdan Kozarzewski

University of Information Technology and Management

ul. H. Sucharskiego 2, 35-225 Rzeszów, Poland.

*E-mail address:*bkozarzewski@mail.wsiz.rzeszow.pl

## Abstract

A method based on mapping a symbolic sequence into a set of patterns (strings resulting from the sequence parsing) is proposed as a tool for the reconstruction of ancestral sequences. The set union of patterns comprises all the patterns present in the family of related proteins sequences of an extant species. The set of most frequent patterns among protein sequences is selected and concatenated. The resulting sequence of amino acids is supposed to be the ancestral protein of the family. No sequences alignment and phylogenetic tree of the species family are necessary. The method is used for inferring the ancestral amino acid sequences of thirteen mitochondria-encoded protein families of mammal species. Statistical distribution of the similarity between extant and ancestral sequences exhibits some structures related to environmental changes in the past.

## 1. Introduction

The majority of methods of DNA or protein sequence analyses rely on previous aligning corresponding sequences. However, alignment algorithms suffer from inherent drawbacks, in particular, for long sequences, see for example [1] and references therein. As an alternative, some pattern matching approaches have been proposed in [2]. Pattern matching approach starts with the mapping of the sequence onto a vector of patterns (strings of different length). The set of patterns results from the specific parsing of the symbolic sequence.

In [2] the algorithm proposed by Ke and Tong, [3] to define a measure of complexity of the binary sequence, after minor modifications was used as a tool for discovering a set of words which are considered as patterns representing a symbolic sequence over an arbitrary alphabet. The set, which will be called a word spectrum, consists of ordered, distinct, non-overlapping words of the size greater than 2 (with the exception of the last word which can be shorter than 3). The parsing algorithm and pattern matching approach together make a new, powerful tool for the symbolic sequence analysis.

Recent accumulation of the nucleotide and amino acid sequence data combined with the computational power have opened a way to realise the idea of Zuckerkandl and Pauling, [4] reconstructing amino acid sequences of ancestral proteins by tracing changes in the sequences of related proteins found in contemporary organisms. The reconstruction of ancestral nucleotide and/or amino acid sequences allows for inferring information about past events in the evolution of species. The computationally reconstructed protein sequence allows testing the sequence in the laboratory by actually resurrecting ancient proteins themselves.

All the ancestral sequence reconstruction methods used so far require as input both multiple sequence alignment of the existing sequences and a corresponding phylogenetic tree. They output a statistical inference of the ancestral sequence at any internal node of the phylogenetic tree. Any uncertainties associated with phylogenetic hypotheses and the methodological issues associated with the inference of ancestral protein sequences can lead to a false reconstruction, [5].

The accuracy of any ancestral reconstruction depends also on the phenomenological models of sequence changes due to the evolutionary process and suffers from possible errors or bias in the reconstruction. Williams *et al.*, [6] performed computational population evolution simulations and compared the thermodynamic properties of the true ancestral sequences with the properties of the sequences inferred by the three mentioned methods. They concluded that the methods may sometimes lead to an incorrect reconstruction of the functional properties of an ancestral sequence.

In the present work a new method for inferring ancestral amino acid sequences, based solely on a set of sequences of related proteins present in an extant species is proposed. No phylogenetic tree or evolutionary model is necessary. The phylogenetic tree can be inferred from a similarity matrix of the set, as a by-product, as shown in [2].The ancestral sequence, which is an approximation of historical reality, depends on the set of extant sequences used. The more ex-tant sequences are included in the set the more accurate would an ancestral sequence be. The preliminary result (unpublished) was obtained by the author when the method was applied to a set of 62 rhodopsin sequences for inferring the amino acid sequence of a hypothetical ancestral rhodopsin protein. The result was compared with the conventionally reconstructed ancestral archosaur rhodopsin by *Chang et al.*, [7]. The similarity between them is 0.64.
In the present paper the method is used for inferring the ancestral amino acid sequences of all thirteen mitochondria-encoded protein families of mammal species. Every ancestral protein is also compared with the modern proteins of the family. Resulting set of similarities may lead to better understanding of the evolutionary processes for specific protein families. For example, statistical distribution of the similarity between extant and ancestral sequences shows some structures related, in my opinion, to environmental changes in the past.

## 2. Methods

2.1.Pattern discovering algorithm

The word spectrum consists of consecutive, distinct, chaotic (not periodic) strings of arbitrary length. Suppose there is a primary sequence $C$ of symbols $c_1, c_2,...,c_n$ . Suppose $S_t$ is a set of words obtained so far and the first symbol of the new word $w$ is $c_i$. The word is formed as a result of a specific procedure of appending symbol $c_i$by the following symbols in three steps.
Step1. String $Q = c_i$ is neither periodic nor chaotic because there is only one symbol in it. So it has to be appended by the next symbol. Appending is continued until some symbol $c_{i+j+l}$ repeats one of the symbols, say $k$ -th, in the string $Q = c_i,...,c_{i+j}$.
Step 2. Let $P = c_k$and $R = c_{i+j+1}$, so far they are equal. Both strings are appended $P = c_k\, c_{k+1}$, $R = c_{i+j+1}c_{i+j+2}$ and so on, until they become different. Then string $Q$ found in Step 1 is appended by string $R$, and the new string is $Q = QR$.
Step 3. Set $S_t$ of words is searched for the presence of string $Q$. If string $Q$ is found, it is appended by the following (next to the last symbol of $Q$) symbol of $C$ becoming $Q = Qc_{i+j+k+1}$. Appending is continued until some string $Qc_{i+j+k+l}$ does not replicate any word from $S_t$. The string $w = Qc_{i+j+k+l}$ becomes the new word of the spectrum representing sequence $C$. It may happen that several last symbols of $C$ cannot be processed by the above replication. They make a new word.
The code of the parsing algorithm is available on request. The result of the sequence decomposition is a set of ordered, distinct and non-overlapping words which will be called the word spectrum of the primary symbolic sequence $C$. The spectrum is a very rich resource of information on the symbolic sequence over arbitrary alphabet.

2.2.Similarity between sequences

Measuring the similarity between symbolic sequences is essential in many data analysis. When spectra $S_1$ and $S_2$ of two sequences $C_1$ and $C_2$are known, the most natural similarity measure can be defined as a set theory intersection of $S_1$ and $S_2$ against the total number of words in both spectra

$$s(C_1,C_2)=\frac{2l(\mathrm{int}(S_1,S_2))}{l(S_1)+l(S_2)}$$

Here $\mathrm{int}(S_1,S_2)$ is a set of words that the two spectra share (set theory intersection of $S_1$ and $S_2$ ), and $l(A)$ means the length (number of words) of set $A$. The value of similarity measure $s(C_1,C_2)$ varies between 0 when the spectra are disjoint sets and 1 when sequence $C_1$ and $C_2$are mutual copies.

2.3.Union set of spectra

The sequence analysis often has to deal with a set $C$ of sequences $C_1$, $C_2$,..., $C_n$. The union set is defined as a set theory union of spectra $S_1$, $S_2$,..., $S_n$ of all sequences $C_1$, $C_2$,..., $C_n$.

$$U = \mathrm{union}(S_1,S_2,...,S_n)\ .$$

The union set includes words from all spectra but with no repetitions. The union set $U$ of words representing set $C$ plays the crucial role in the reconstruction of the ancestral sequence.

2.4. Algorithm for reconstruction of ancestral sequence

The present day proteins result from a long evolution of ancient life forms. Their proteins do not survive long lasting destructive processes. Nevertheless, ancient proteins can be studied by inferring ancestral sequences by means of appropriate mathematical techniques. Reconstructed ancestral protein sequences can help understand the evolutionary processes and mechanisms by which proteins acquired their present functions, [8]. The fundamental assumption of the present approach is that spectra of similar sequences and the spectrum of a common ancestral sequence share a significant number of words. For example, a given protein family (a set of amino acid sequences) of extant species predicts with some probability the ancestral amino acid sequence of the family. Let $C = C_1$, $C_2$,..., $C_n$ be a set of extant sequences. The algorithm for building an ancestral sequence of set $C$ consists of the following steps.

Step 1. Find word spectra $S_1$, $S_2$,..., $S_n$ of all sequences.
Step 2. Generate the union set of the spectra, suppose it is the set of words $U = w_1, w_2,...,w_N$. It is convenient to represent the set as a column $N$ vector.
Step 3. Determine intersection of spectrum $S_1$ with the union set. It is a set of words the spectrum shares with $U$. It is convenient to represent the set as a numeric column $N$ vector. Its $i$-th nonzero component is equal to the index of word $w_i$ in spectrum $S_1$. The vectors corresponding to all the spectra form the columns of a matrix of size $N \times n$. The vector of union words from Step 2 appended by $N \times n$ matrix from Step 3 above form the $N \times (n+1)$ table $W$, that comprises all available data on the words present in the set of $n$ symbolic sequences.
Step 4. Find the conservation number (it says how many spectra possess this particular word, cn in short), mode and frequency of elements in each row of table $W$. The final result is a table $A$ consisting of $N$ rows and 4 columns.
Step 5. Select a word of the highest frequency from the group of words of mode 1. There may be several such a words, then select one of highest conservation number. Make the word and its attributes (cn, mode and frequency) the first row of a table $B$ of 4 columns. Continue for words of mode 2, 3, .., until the frequency becomes lower than approximately $0.1n$. Order the rows of table $B$ according to the increasing mode.
Step 6. All the words in the first column of table $B$ make a set of potential segments of the ancestral sequence of set $C$ of extant sequences. In the simplest case when there is only one word of the highest conservation number and frequency, for every mode, concatenate the words in the ascending order of the mode. The resulting string of symbols is supposed to be the ancestral sequence. Otherwise, one has to select one word from every group of words of the same mode. Concatenate the words in the ascending order of the mode. The resulting string of symbols is the variant of ancestral sequence.

## 3. Results

The mitochondria coding sequences of $N_s = 360$ mammal species were downloaded from NCBI, http://www.ncbi.nih.gov/. Their accession numbers and species names are listed in the Supplementary material (available on request), file Table_1. For each family, the similarity between extant sequences and ancestral sequence were calculated. When the similarities are considered as instances of a random variable, the statistical properties of the variable, like distribution function of a member of family proteins can be defined. In the following, results for four families are presented and distribution function of similarity between ancestral sequence and descendant sequences for each family is discussed.

3.1. COX1 gene coding protein family

According to Step 1 of the algorithm, the spectra (each including about 80 words) of all the sequences were obtained. For example, the first three words of the protein spectrum of *Ovis canadensis* species are: 'MFINRWLFS', 'TNHKDIGTL', 'YLLF'. Step 2, union set $U$ of the spectra consisting of 1206 words was found. Step 3, intersection of spectrum $S_1$ (of *Acomys cahirinus* species) with the union set was found; it is a column vector with 80 nonempty elements out of 1206. The first nonempty element is in the 16-th row corresponding to word 'AGAS' (which is the 16-th word of $U$) and its value 22 means that the word is 22nd in the spectrum $S_1$. The other 359 columns have a similar interpretation. Step 4, the second word of union set $U$ is 'AAV'. The word is present in only one spectrum (its conservation number is 1), which comes from the 254-th column representing the extant sequence (*Pipistrellus abramus* species) and is 74-th word in the spectrum. Therefore, in the second row, the second, third and fourth columns of table $W$ there are conservation number, cn = 1, mode = 74 and frequency = 1. Similarly, in the 1143-th row (corresponding to the word 'WFFG', of highest conservation number) the next three columns are cn = 360, mode = 38 and frequency = 210. The selection in Step 5 leads to a relatively small table of 82 rows, and 5 columns, as shown in Table_2 of the Supplementary material. Then, according to Step 6 the ancestral sequence results from consecutive concatenation of all the words from the first column of the table. The ancestral sequences of amino acids are given in Appendix and the file 'ancestral_sequences.txt' in the Supplementary material.
To compare the ancestral sequence with extant sequences, the spectrum $S_{anc}$ of the sequence is necessary. Then $N_s$ similarities between ancestral and extant sequences were calculated and ordered according to the increasing similarity. If the similarity to the ancestral sequence is considered as a random number, then index $N(s)$ of extant sequence of similarity $s$ is interpreted as the number of sequences (from among the set of $N_s$ sequences) of similarity less or equal to $s$. $N(s)/N_s$ means nothing else but an experimental cumulative distribution function (*ecdf*) of random variable $s$. From Fig. 1 it follows that in the wide range of similarities, *ecdf* can be roughly approximated by a linear function. Therefore $N(s)/N_s$ in the first approximation can be represented by the cumulative distribution function (*cdf*) of uniform distribution, as shown in Fig.1. The partial distribution function ($pdf$) is constant = 2.63 in the range $\Delta s$ = (0.56, 0.94) of similarity and zero elsewhere. It means that 360 sequences are distributed uniformly with density $\Delta N(s)/\Delta s = 2.63 * N_s$ = 945 per unit similarity measure within the range $\Delta s$, and 0 elsewhere.

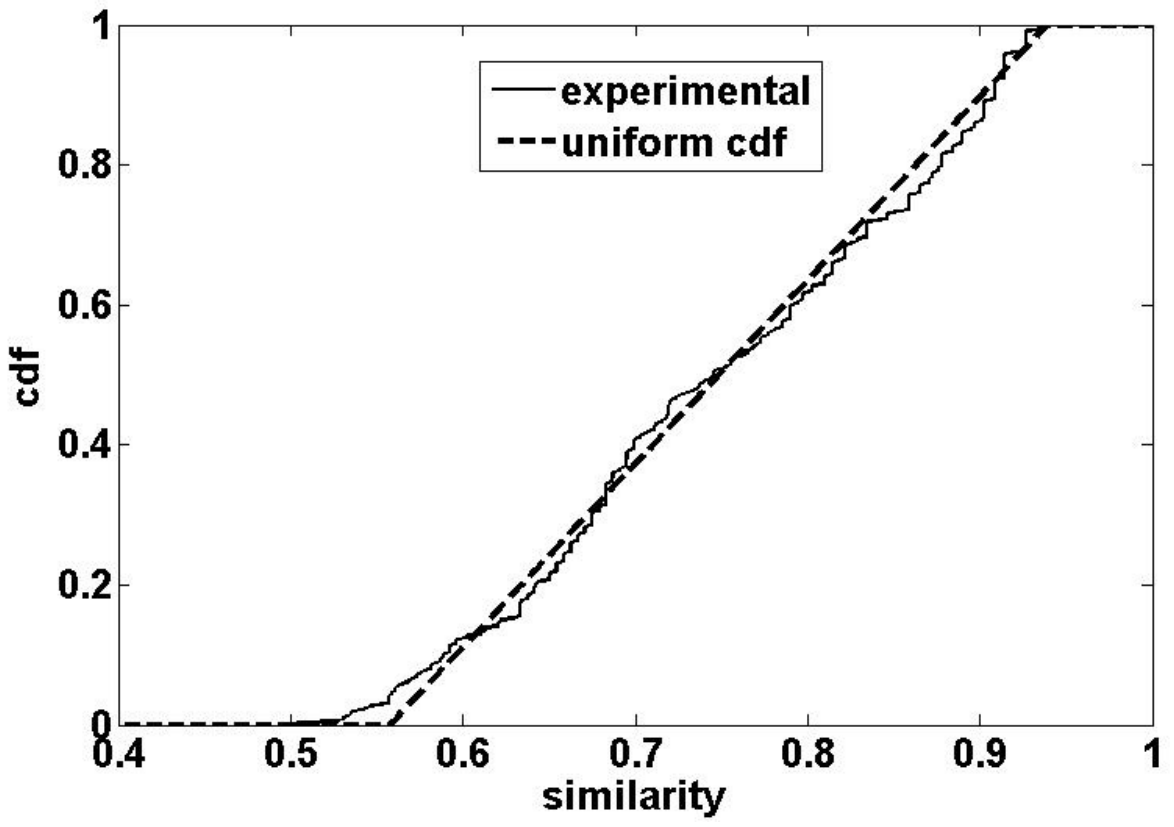

Fig.1. *Cdf* of COX1 gene encoded protein family and its uniform approximation

After Zuckerkandl and Pauling's [4] paper it is generally accepted that the time since divergence of the species bearing a protein sequence is related to the similarity between the sequence and ancestral sequence of the family *via* some function $f(s)$. The sequence of lowest similarity $s_{\min}$ comes from the latest extant descendant of the ancestral sequence, $s_{\min}$ corresponds to the present time ($t$ = 0). The sequence of highest similarity $s_{\max}$ belongs to the first not extinct species of the family and $s$ = 1 corresponds to the time of ancestral sequence divergence. Therefore the divergence time of the sequence of similarity $s$ is given by $t = f(s) - f(s_{\min})$. Time difference $\Delta t = f(s_{\max}) - f(s_{\min})$ is a measure of the lifetime of sequence $C$. If, according to the molecular clock hypothesis $f(s)$ is a linear function, then $t = -\alpha s - \beta$ , and divergence time $\Delta t = \alpha(s_{\max} - s_{\min})$ . Assuming several calibration points one can translate the similarity between the extant and ancestral sequences into absolute geological times. If the above reasoning is correct, then the linear dependence of *ecdf* on similarity is equivalent to the molecular clock hypothesis. The reliability of molecular clock methods depends on the accuracy with which the genetic similarity (or distance) is estimated, and on the appropriateness of the calibration [9]. A more detail inspection of experimental *cdf* shows minor, not random, deviations from uniform distribution. Much better (than linear) fit to the experimental distribution is obtained with a piecewise function of four logistic distributions

$$cdf(s) = b + \frac{a}{1 + e^{\frac{s-m}{\sigma}}}$$

where the constant $1/\sigma$ defines the growth rate and $a$ is the carrying capacity (limiting the population size).
The set of parameters of logistic distribution and the range of applicability of each of them are given in Table 1.

**Table 1.** Parameters and ranges of applicability for logistic distributions

| $m$ | $\sigma$ | $a$ | $b$ | above $s$ |
|---|---|---|---|---|
| 0.580 | 0.022 | 0.170 | 0 | 0.494 |
| 0.672 | 0.00 | 0.320 | 0.140 | 0.634 |
| 0.800 | 0.041 | 0.416 | 0.420 | 0.720 |
| 0.898 | 0.020 | 0.320 | 0.715 | 0.861 |

Once theoretical *cdf* is known, the partial distribution function is easy to find

$$pdf(s) = \frac{d}{ds} cdf(s) .$$

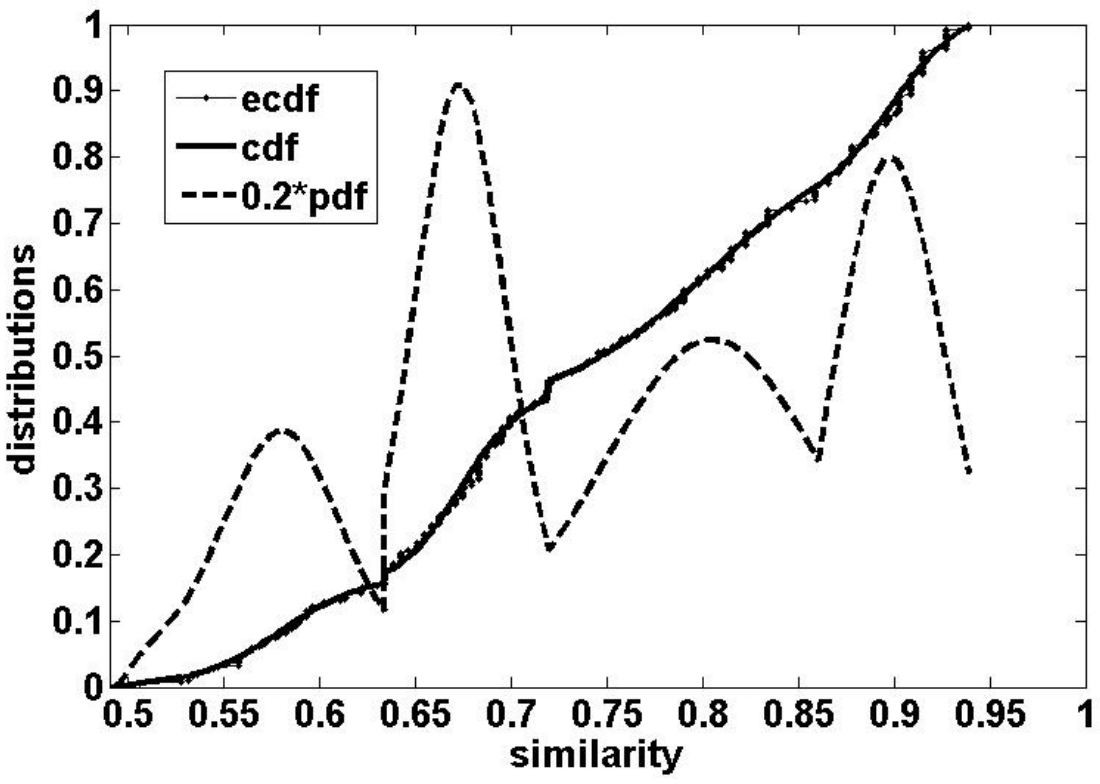

Fig.2. Distribution functions of COX1 encoded protein family

Fig.2 shows plots of three distribution functions of similarity between the ancestral sequence and extant sequences: experimental cumulative distribution function, theoretical (fitted) *cdf* and *pdf*. The density of protein sequences $\Delta N(s)/\Delta s = N_s pdf(s)$ varies significant across the similarity range. There are no sequences of similarity greater than 0.94, which probably means that they have become extinct by now, there are no sequences of similarity and less than 0.49 either, probably they have not diverged yet.

The logistic function of time is often used as a model of population growth. The parameters $1/\sigma$ and $a$ depend (among others) on external conditions. After change in characteristics of the environment, the population size and/or growth rate adjusts to the change. One can make hypothesise that the variation of parameters of the logistic distribution function reflects past incidents of relatively drastic changes in environmental conditions.

3.2. NADH1 gene coding protein family

Following the steps like those in 3.1 the spectra of sequences, union set of 2436 words long, ancestral sequence and similarity between each of 360 protein sequences and ancestral sequence were calculated. This time, the experimental cumulative distribution function of similarity can be quite well fitted to the superposition of three logistic functions. The derivative of theoretical *cdf* with respect to similarity yields a partial distribution function. The set of parameters and ranges of their applicability for each them are given in Table 2.

**Table 2.** Parameters and ranges of applicability for logistic distributions

| $m$ | $\sigma$ | $a$ | $b$ | above $s$ |
|---|---|---|---|---|
| 0.190 | 0.032 | 0.450 | 0.025 | 0.000 |
| 0.465 | 0.041 | 0.320 | 0.460 | 0.317 |
| 0.805 | 0.032 | 0.245 | 0.775 | 0.654 |

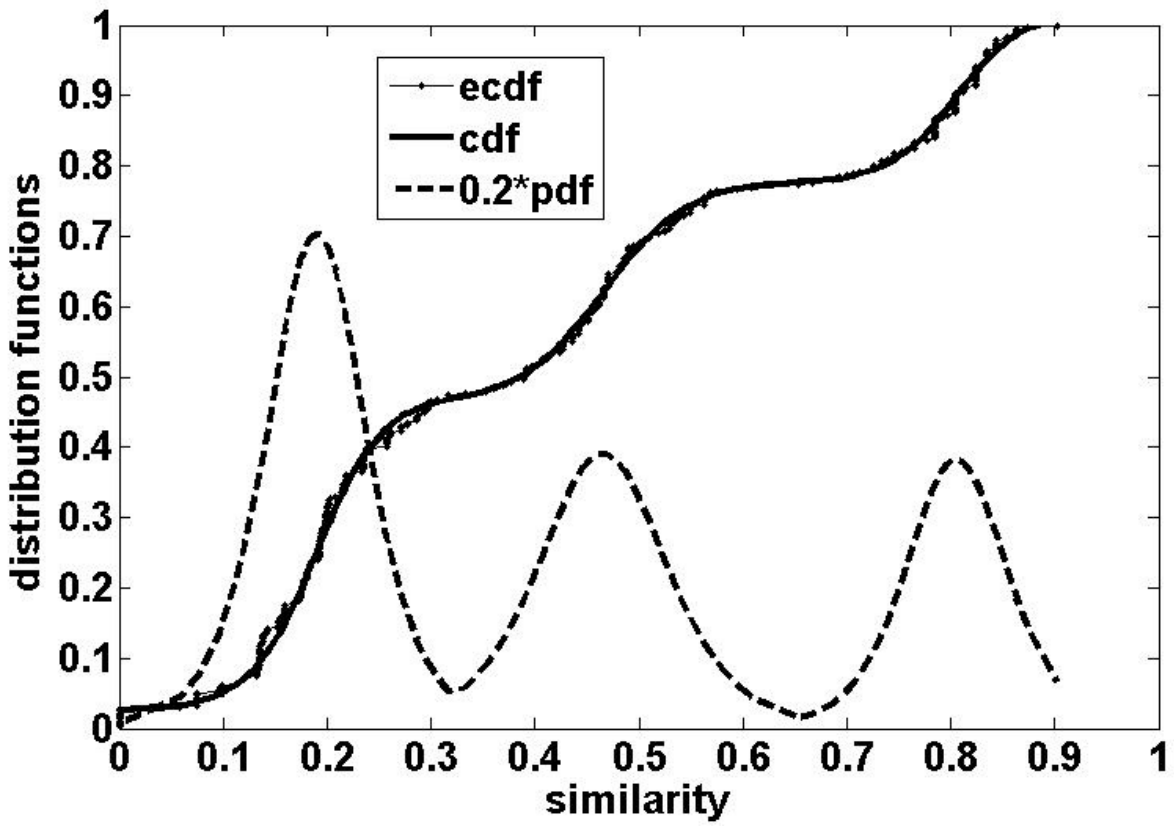

**Fig.3.**Distribution functions of NADH1 gene encoded protein family

Fig.3 shows the plots of the three distribution functions: experimental cumulative distribution function, theoretical *cdf* and *pdf*. The maximums of *pdf* at similarity around 0.2 and 0.45 and 0.8 means that in the past there were at least three periods of a relatively high population of species bearing dehydrogenase subunit 1 protein. The latest extant descendants of the ancestral sequences are quite different (of zero similarity) from the ancestral sequence, there are ten of them.

3.3. NADH6 gene coding protein family

Only 358 sequences of NADH6 mammal protein have been found, their spectra were in average 30 words long. The union set $U$ of spectra consists of 2269 words. The ancestral sequence is given in the file 'ancestral_sequences.txt' in the Supplementary material. As before, the relative number of sequences of similarity (to the ancestral sequence) less than or equal to $s$, $N(s)/N(358)$ is interpreted as an experimental cumulative distribution function of random variable $s$. Taking into account the most pronounced features of *ecdf* function the piecewise distribution of two logistic cumulative distribution functions is enough to approximate *ecdf*. Their parameters and rages of applicability are shown in Table 3.

**Table 3.** Parameters and ranges of applicability for logistic distributions

| $m$ | $\sigma$ | $a$ | $b$ | above $s$ |
|---|---|---|---|---|
| 0.072 | 0.023 | 0.470 | 0.010 | 0.000 |
| 0.425 | 0.069 | 0.554 | 0.460 | 0.168 |

The plot shows all distribution functions, *ecdf* and averaged *cdf* and 0.2*pdf*.

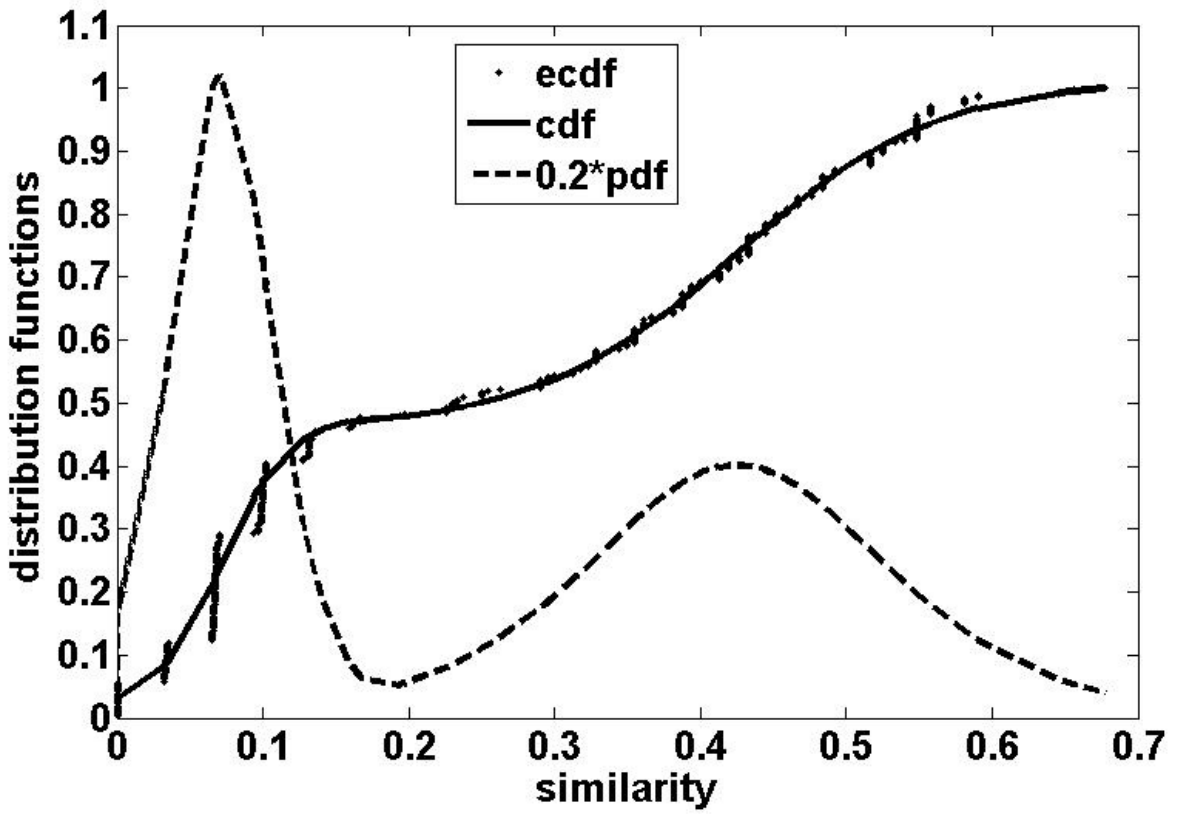

Fig.4. Distribution functions of NADH6 encoded protein family

In Fig.4 the plots of experimental – *ecdf* and theoretical distribution functions *cdf* and 0.2*pdf* are shown. The latest 20 extant descendants of the ancestral sequences are quite different (of zero similarity) from the ancestral sequence. The range of similarity between 0.0 and 0.2 is interesting because there are no extant proteins in several intervals of 0.03 in width on the similarity axis. Instead, the tenths of sequences are densely packed within the interval of few $10^{-3}$ width. The probability density to find an extant sequence of similarity to ancestral sequence exhibits very deep minimum around 0.186. If our set of 358 proteins is representative of all extant mammal species, then there are relatively few species bearing the protein sequence of similarity around 0.19.

## 4. Discussion

The question arises how reliable are the inferred ancestral sequences. The direct comparison with true ancestral sequences is not possible. An indirect answer comes from the following reasoning. We restrict ourselves to COX1 protein family. The set *S* of 360 extant sequences is ordered according to increasing similarity to inferred ancestral sequence. The next one in the row is the inferred ancestral sequence. Then let us consider the subset consisting of *n* first sequence of the set *S*. Applying the method presented above we can find inferred ancestral sequence of the subset. The similarity between the ancestral sequence and the *n* +1 sequence in *S* is easy to find. For *n* = 357 the next sequence comes from *Capra sibirica* species. The similarity between the ancestral sequence of the subset and 358-th sequence is 0.94. One can say that the reconstructed ancestral sequence of the subset and the true predecessor sequence of the subset are similar in 94%. For *n* = 346 the next sequence comes from *Axis porcinus* species. The corresponding similarity is 0.92. The average similarity for 20 randomly selected subsets is 0.76 with minimum of 0.53 for subset of *n* = 101. In general the longer subset the closer is inferred sequence to the *n* +1 sequence (hypothetical true ancestral sequence of the subset).

The discontinuity in distribution of extant sequences along the similarity axis is present in several protein families. It is particularly well visible in the case of ATP8 gene coded protein family. It follows from Fig. 5 that there are groups of sequences of close similarities to the ancestral sequence, separated by wide (of 0.08) intervals of similarity value not allowed for proteins (see inset as well). The groups of proteins of the same or very close similarity between extant descendants and the ancestral sequences consists of 61 species (similarity 0), 170 (similarity around 0.1). To get some insight into such behaviour some simulation was done. Starting from some real coding nuclear sequence *C* considered as ancestral and simulating random mutation in consecutive sequences, an artificial family of coding sequences was obtained. Then all members of the family were converted into amino acid sequences, using the vertebrate mitochondria genetic code.

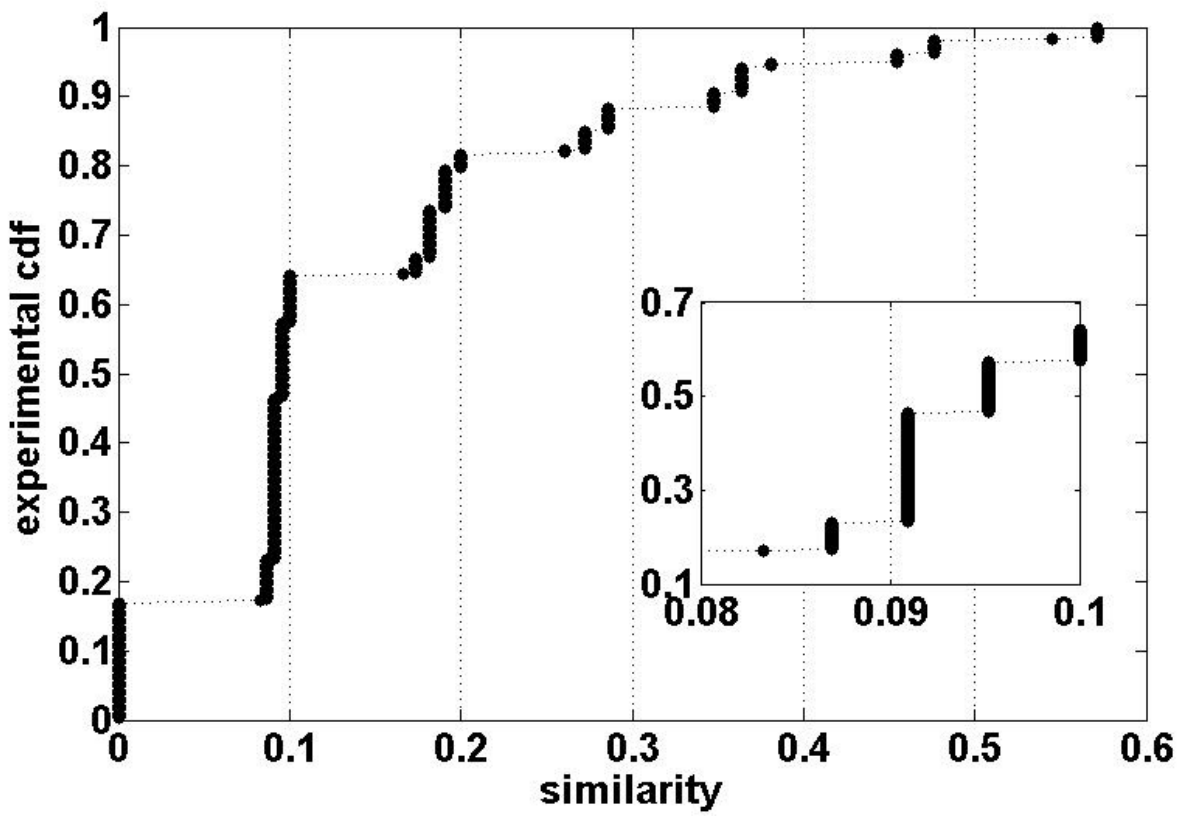

**Fig.5.** Experimental *cdf* of ATP8 encoded protein family

The set of similarities between the ancestral (corresponding to *C*) protein and its descendants was analysed under different substitution rates between sites. Experimental *cdf*, characteristic of ATP8 encoded family is observed only for low substitution rate.

To determine if the computational reconstructed ancestral protein would be functionally active and thus could be regarded as a reasonable approximation of a true ancestral sequence, the corresponding genes have to be synthesised and then expressed in a mammalian cell line in tissue culture.

## References

[1] S. Kumar, A. Filipski, *Multiple sequence alignment: In pursuit of homologous DNA positions*. Genome Research, **17**, 127-135 (2007).

[2] B. Kozarzewski, *A method for nucleotide sequence analysis*, Computational Methods in Science and Technol ogy, **18**, 5-10 (2012).

[3] D.G. Ke, Q.Y. Tong, *Easily adaptable complexity measure for finite time series,* Phys. Rev., **E77**, 066215:23 (2008).

[4] E. Zuckerkandl, L. Pauling, *Molecular restoration studies of extinct forms of life,* Acta chemica scandinavica, 17, 9-16 (1963).

[5] D. Schluter, *Uncertainty in ancient phylogenies*, Nature **377**, 108–110 (1995).

[6] P.D. Williams, D.D., Pollock, B. P. Blackburne, R. A. Goldstein *Assessing the accuracy of ancestral protein reconstruction methods*, PLoS *Comput. Biol.* 2, 598-605 DOI-10.1371/journal.pcbi.0020069 (2006).

**[7]** B.S.W. Chang, K. Jönsson , M. A. Kazmi, M.J. Donoghue, T. P. Sakmar *Recreating a functional ancestral ar-*

*chosaur visual pigment*, Mol. Biol. Evol. 19, 1483–1489 (2002).
[8] M.J. Harms, J.W. Thornton, *Analyzing protein structure and function using ancestral gene reconstruction*, Current Opinion in Chemical Biology, **20**, 360-366 (2010).
[9] L. Bromham, D. Penny, *The modern molecular clock*, Nature Reviews Genetics 4, 216-224 doi-10.1038/nrg1020 (2003).

## Appendix

Mitochondria-encoded mammal ancestral proteins, synthetic construct

>ATP synthase F0 subunit 6
MNENLFASFITPTMMGLPIVILIIMFPSILFPSSRLISLQQWMLQLQWLVQLIHNTKGQTWTLMLMSLIL
FIGSTNLLGLLPHSFTPTTQLSMNLGMAIPLWAGAVITGFRNKTKASLAHSLAHFLPPLIPMIPMLVIIT
ISLFIQPVALAVRLTANITAGHLLIHLIGGATLALMSISTSISPIILLILLTILEFAVAMIQAYVFTLLVSLYLHDNT

>ATP synthase F0 subunit 8
MPQLDTSTWLTMILSMFLVLFIIFQLKISKHNFYHNPNFHFNWESKWTKIYSPLSLLPLLPQQS

>NADH dehydrogenase subunit 1, ancestral sequence
MFMINILMLIIPILLAVAFLTLVERKVLGYMQLRPNVVGPYGLLQPIADAIKLFTKEPLRPATSSISMFI
LAPILALSLALTMWIPLPMPYPLINMNLGVLFMLAMSSLAVYSILWSGWASNSKYALIGALRAVAQTISY
EVTLAIILLSVLLMNGSFTLSTLIITQEQVWLIFPAWPLAMMWFISTLAETNRAPFDLTEGESELVSGFN
VEYAAGPFALFFMAEYANIIMMNIFTTTLFLGAFHAFHNPYMPELYTINFTIKSLLLTISFLWIRASYPR
FRYDQLMHLLWKNFLPLTLALCMWHVSLPILLSSIPT

>cytochrome c oxidase subunit 1
MFINRWLFSTNHKDIGTLYLLFGAWAGMVGTALSLLIRAELGQPGTLLGDDQIYNVIVTAHAFVMIFFMV
MPIMIGGFGNWLVPLMIGAPDMAFPRMNNMSFWLLPPSFLLLLASSMVEAGAGTGWTVYPPLAGNLAH
AGASVDLTIFSLHLAGVSSILGAINFITTIINMKPPAMSQYQTPLFVWSVLITAVLLLLSLPVLAAGITMLL
TDRNLNTTFFDPAGGGDPILYQHLFWFFGHPEVYILILPGFGMISHIVTYYSGKKEPFGYMGMVWAMMSI
GFLGFIVWAHHMFTVGMDVDTRAYFTSATMIIAIPTGVKVFSWLATLHGGNIKWSPAMLWALGFIFLFTV
GGLTGIVLANSSLDIVLHDTYYVVAHFHYVLSMGAVFAIMGGFVHWFPLFSGYTLNDTWAKIHFAIFVGV
NMTFFPQHFLGLSGMPRRYSDYPDAYTTWSSMGGSFISLTAVMLMIFIIWEAFASKREVLTVDLTTTNLE
WLNGCPPPYHTFEEPTYVNLK

>cytochrome c oxidase subunit 2
MAYPMQLGFQDATSPIMEELLHFHDHTLMIVFLISSLVLYIISLMLTTKLTHTSTMDAQEVETIWTILPA
IILILIALPSLRILYMMDEINNPSLTVKTMYEYTDYEDLNFDSYMIPTQELKPGSFDSYLLEVDNRVVEV
DNRVVLPMEMTIRMLISSEDVLHSWAVPSLGLKTDAIPGRLNQTTQCSEICGYGQCSEIGSNHSFMPIV
LELVPLKYFEKWSASML

>cytochrome c oxidase subunit 3
MTHQTHAYHMVNPSPWPLTGALSALLMTSGLIMWFHFNSTTLLGLLTNMLTMYQWWRDIIRESTFQGHHT
PTVQKGLRYGMILFIISEVLFFTGFFWAFYHSSLAPTPELGGCWPPTGIHPLNPLEVPLLNTSVLLASGV
SITWAHHSLMEGNRNHMLQALFITIALGVYFTLLQASEYYEAPFTISDGVYGSTFFMATGFHGLHVIIGS
TFLIVCFFRQLKFHFTSSHGFEAAAWHFGFEAAAWYWHFVDVVWLFLYVSIYWWGS

>cytochrome b
MTNIRKTHPLMKIVNNAFIDLPAPSNISSWWNFGSLLGICLILQILTGLFIITTAFSSVTHICRDVNYGW
IIRYMHANGASMFFIRYLHANGASVGRGLYYGNIGVILLLFTVMATAFMGYVLPWGQMSFWGATVITNLL
SAIPYIGTNLVEWIWGGFSVDKATLTRFFAGGFFHFIALAMVHLLFLHETGSNNPHETGSNNPSDKIPFH
PYYTIKDILGALLLIYTIKDLLGLLVLFSPDLLGDPDNYTPANPLNTPPHSTPPHYAILRSIPNKLGGVL
ALVLSILIYAILRSVPNKLGPLLHTSKQRSMMFRPFSQCLFWILVTLTWLTWIGGQPVEHPYIIIGQLAS
ILYFLLILVLMLKWENNLKW

>NADH dehydrogenase subunit 1
MFMINILMLIIPILLAVAFLTLVERKVLGYMQLRPNVVGPYGLLQPIADAIKLFTKEPLRPATSSISMFI
LAPILALSLALTMWIPLPMPYPLINMNLGVLFMLAMSSLAVYSILWSGWASNSKYALIGALRAVAQTISY
EVTLAIILLSVLLMNGSFTLSTLIITQEQVWLIFPAWPLAMMWFISTLAETNRAPFDLTEGESELVSGFN
VEYAAGPFALFFMAEYANIIMMNIFTTTLFLGAFHAFHNPYMPELYTINFTIKSLLLTISFLWIRASYPR
FRYDQLMHLLWKNFLPLTLALCMWHVSLPILLSSIPT

>NADH dehydrogenase subunit 2
MNPIIFIIILITGTLLSSHMISSHWLLIWIGFEMNMLAIIPIMMKTKYFLTQATASSTASMLLMMAVIIN
LMFSGQWTVMKLFNPVASMFMTMALTMKPFHFWVPEVTQGIPLSSGLILLTWQKLAPMSVLYQISPSILN
MILTISILLTLSSIMIGGGLLNQTQLNQTQLRKIMAYSSIAHMGWMTAVLPYNPTMTLLNLIIYIIMTST
MFTLFMANSTTTTLSLSHTWNKMPVMTSTLLSTWNKTPLGGLMGGLPPLSGFMPKWMIIQELTKNDSLIP
TFMAITAMMALLNLYLYFYMFYMRLTYSTALTMKMKWFPSTNNMQMTLLPQFSTTKSTMLLPLTPILSIL

EPLTPISSLLSILEVLD

>NADH dehydrogenase subunit 3
MNLMLALLTNFTLASLLVIIAFWLPQLNVYSEKTSPYECGFDPMGSARLPFSMKFFLVAITFLLFDLEIA
LLLPLPWASQTTNMALLLILFLISLLAASLAYWTEGLEWTEK

>NADH dehydrogenase subunit 4L
MSLVYMNIMMALLGMYRSHLMSYRSHLMSSLLCLEGMMLILNSHFTLASMMPFTLANMAPIILLVFAACE
AALGLSLLVMVSNTYGTDYVQNLNLLQC

>NADH dehydrogenase subunit 4
MLKYIIPTMMLMPLTWLSKNNMIWINSLLISLTSLLLMNQFSDNSQFNDNSFFSDSLSTPLLILTMWLLL
LPSQHHLSQSHKKLLLQLFLIMTFTATELIMFYILILFEATLVTELILFPTLIIITRWGNQTERLNAGLY
FLFYTLVGSLPLLVSLPLLNITGSLNFLVLQLQYWVQPYWVQPLSNSWSNVFMWLACMMAFMVKMPLAC
MMALYGLHLWLPKAHVEAPIAGSMVLAAILLKLGGYLNPLTGMMRFMAYPFIITILLWGMIMTSSICLRQT
DLKSLIAYSSVSHMALVIVAILIQTPWSYMGATALMIAHGLTSSMLFCLANSNYERIHSRTMILARGLQE
RTHSRIERTHSRTMTLLPLASLTNLALPPTINLIGSLTNLALPPTINLIGELFVVMMVITALYSLYMITA
LYSLYMRGKYTHHILITTQSPSFTRENALMSLHMLPLLLLLSLNPKIILGPLYNPKIILGPIYLNPKVILG

>NADH dehydrogenase subunit 5
MNMFSSFLLTIPIMTTSYPQYVKTTISYAFITSMIPTMMFSNWHWMKLSLSFFKLDYFSKMDYFSMMFVP
VALFVTWSIMEFSMWYMHSDPNINQFFKYLLLFFQLFIGWEGVGIMSFLLIGWWYGRTDANTAALQAILY
NRIGDIGFTAALQAILYNRIGDIGFILAMAWFLTNLNTWDLQQIGKSAQFGLKSAQFGLHPWLPSAMEGP
TPVSALLHSSTVVAGIFLMIRMVVAGIFLLITLCLGRFYPLTENNKFTAICALLGAITTLTQNDIKKIIA
FSTSSQLGLMIVTIGMVTIGINICTHAFFKAMLFMCSGSIEQDIRKMGGLIHSLNDEQDIRKMGGLFKAM
PFTTTAYSKDLIIESLALTGMPFLTGFLITLIAAANAWALLMTLIATSFTAIYSTRIIENNPFFALLGQP
RFPTGSIFAGFPQMTMPFLMNSIKRLMIGSLFAGFIISIISNNIPPTTIPQLTMPYYLKMMALTVTINLL
GLEISNMTQNLFKFSNMLGYFPTIMHSQKSALIWLELLDLDLIYFLSFLIKTTSLIQTKMSIMLYFLSVT
NQKGLIKLYFLSFLVLIQMKMSMILFNFHE

>NADH dehydrogenase subunit 6
MMTYIVFILSIIFFVGFSSKPSPIYGGLYGGLVGCGIGCGIFLGLMVFLIYLGGMLVVFGYTTAVFGYTT
AMATEQYPEIWVSNKTVLGAFVTGLLMEFLMVYYVLKDEEVVEIVFGLGDWVIYDTGDSGFFSEEAMGIA
ALYSYGTWLVIVTGWSLLIGVVVIMEITRGN